\documentclass[aps,prd,preprint,superscriptaddress,showpacs]{revtex4-1}

\usepackage{amsmath, amsfonts, amsthm, amssymb, mathrsfs, enumerate, graphicx}

\newcommand{\sech}{\mathrm{sech} \,}
\newcommand{\R}{{\mathbb R}}

\begin{document}


\title{Curved traversable wormholes in (3+1)-dimensional spacetime}


\author{Vee-Liem Saw}
\email[]{VeeLiem@ntu.edu.sg}
\affiliation{Division of Physics and Applied Physics, School of Physical and
Mathematical Sciences, Nanyang Technological University, 21
Nanyang Link, Singapore 637371}

\author{Lock Yue Chew}
\email[]{lockyue@ntu.edu.sg}
\affiliation{Division of Physics and Applied Physics, School of Physical and
Mathematical Sciences, Nanyang Technological University, 21
Nanyang Link, Singapore 637371}


\date{\today}

\begin{abstract}
We present the general method of constructing curved traversable wormholes in (3+1)-d spacetime and proceed to thoroughly discuss the physics of a zero tidal force metric without cross-terms. The (3+1)-d solution is compared with the recently studied lower-dimensional counterpart, where we identify that the much richer physics - involving pressures and shear forces of the mass-energy fluid supporting the former - is attributed to the mixing of all three spatial coordinates. Our (3+1)-d universe is the lowest dimension where such nontrivial terms appear. An explicit example, the static zero tidal force (3+1)-d catenary wormhole is analysed and we show the existence of a geodesic through it supported locally by non-exotic matter, similar to the (2+1)-d version. A key difference is that positive mass-energy is used to support the entire (3+1)-d catenary wormhole, though violation of the null energy condition in certain regions is inevitable. This general approach of first constructing the geometry of the spacetime and then using the field equations to determine the physics to support it has the potential to discover new solutions in general relativity or to generalise existing ones. For instance, the metric of a time-evolving inflationary wormhole with a conformal factor can actually be geometrically constructed using our method.
\end{abstract}


\maketitle


\section{Introduction}
\label{sec1}

The first study on curved traversable wormholes was recently explored in \cite{Vee2012}, generalising the research on traversable wormholes to the non-spherically symmetric regime. The preliminary discussion was carried out in (2+1)-d spacetime, where the general method of construction was thoroughly developed. The lower-dimensional spacetime provided a convenient platform for visualisation as well as minimising the heavy algebra due to the lack of spherical symmetry. It was shown that the absence of tidal force resulted in the Einstein field equations reducing to only one equation where the mass-energy density is directly proportional to the Ricci scalar, with no radial tension nor lateral pressure. The key result was the existence of safe geodesics through (2+1)-d curved traversable wormholes which are supported locally by non-exotic matter. This general method is in fact directly applicable to higher dimensions, and was originally conceived when trying to produce a fractal curve by iterative helicalisations on a given smooth curve \cite{Vee2013b}.

Here, we proceed further to study curved traversable wormholes in (3+1)-d spacetime. The general method of construction by adding 2-spheres along a curve is described, with the goal of deriving the (3+1)-d version of the catenary wormhole and the field equations that govern the physics. One may also apply the analysis to the (3+1)-d helical wormhole, though the presence of two cross-terms (see Section 5 in \cite{Vee2012}) would give rise to many extra terms. As the catenary is a plane curve, its metric has no cross-terms (see the lower-dimensional version in \cite{Vee2012}, the (3+1)-d version in the next section) and so the calculations are greatly simplified. The main purpose of this paper is to show that a well-designed curved traversable wormhole permits safe geodesics through it that are locally supported by ordinary matter, avoiding the need for travellers to come into direct contact with exotic matter. This is crucial as the ramifications of ordinary matter meeting exotic matter are unknown.

As the catenary wormhole is a simple example of a non-spherically symmetric spacetime that represents a curved wormhole, it is instructive to compare and contrast it with the spherically symmetric ones. The results of the discussion would provide deeper insights on the geometry of such spacetimes. The layout of this paper is hence as follows: Section 2 pins down the general method for building 3-manifolds of revolution around a given smooth curve, followed by the computation of the relevant tensors for the physical interpretations through the Einstein field equations as well as comparison with those for (2+1)-d, in Section 3. Section 4 focuses exclusively on the (3+1)-d catenary wormhole. The advantages and potential of this geometrical approach in finding new solutions in general relativity form the discussion of Section 5. In that section, we show explicitly how this geometrical technique yields the derivation of the metric for an inflationary wormhole with a conformal factor, which was taken to be an ansatz in \cite{inf,inf2,inf3}. We then conclude this paper in Section 6. A proof that 3-manifolds of revolution generated around plane curves (where the three orthonormal vectors are perpendicular to the tangent vector) have no cross-terms is presented in the appendix.

\section{A general method for generating 3-manifolds of revolution around a given smooth curve}
\label{sec2}

The general method for constructing surfaces of revolution around a given smooth curve has been well-developed in \cite{Vee2012}. Extension of that method to higher dimensions has also been briefly described there, with an explicit construction of the (3+1)-d helical wormhole being illustrated as well. Essentially for our (3+1)-d universe, given a smooth curve that is embedded in a 4-d Euclidean space with parametric equations $\vec{\psi}(v)=(\alpha(v),\beta(v),\gamma(v),\delta(v))$, the 3-manifold of revolution can be generated by adding the oscillatory terms $Z(v)\cos{u}$, $Z(v)\sin{u}\cos{w}$ and $Z(v)\sin{u}\sin{w}$ along three orthonormal vectors $\vec{n}_1(v)$, $\vec{n}_2(v)$ and $\vec{n}_3(v)$ respectively, so the resulting 3-manifold of revolution is
\begin{eqnarray}\label{eq1}
\vec{\sigma}(u,v,w)&=&\vec{\psi}(v)+Z(v)\cos{u}\ \vec{n}_1(v)+Z(v)\sin{u}\cos{w}\ \vec{n}_2(v)\nonumber\\&\ &+\ Z(v)\sin{u}\sin{w}\ \vec{n}_3(v).
\end{eqnarray}
The three oscillatory terms are solutions to the 2-sphere of radius $Z(v)$, just like the two oscillatory terms $Z(v)\cos{u}$ and $Z(v)\sin{u}$ used in generating the surface of revolution are solutions to the 1-sphere of radius $Z(v)$. The radial function $Z(v)>0$ describes the variation of the radius of the 3-manifold of revolution along the given smooth curve, and the three mutually orthonormal vectors $\vec{n}_1(v)$, $\vec{n}_2(v)$ and $\vec{n}_3(v)$ need not be perpendicular to the unit tangent vector of the curve. The constraint $Z(v)>0$ prevents singularities, but it has to be defined with care to prevent possible self-intersections of the resulting 3-manifold. In the (2+1)-d version, a continuous set of circles (1-spheres) is added along the given smooth curve to generate the wormhole surface, whereas here a continuous set of 2-spheres is added to yield the wormhole 3-manifold. Visualising the surface of revolution is relatively easy since it can be embedded into $\R^3$ for our convenience, though trying to do something similar for the 3-manifold would be difficult.

The metric of this 3-manifold can be computed with $g_{ij}(u,v,w)=\vec{\sigma}_i\cdot\vec{\sigma}_j$, where $i,j\in\{u,v,w\}$. The notation $\vec{\sigma}_i$ denotes partial derivative with respect to that variable, for example $\vec{\sigma}_u:=\partial\vec{\sigma}/\partial u$. The (3+1)-d spacetime metric corresponding to such a 3-manifold would be
\begin{eqnarray}
ds^2&=&-e^{2\Phi(u,v,w)}c^2dt^2+g_{uu}(u,v,w)\ du^2+g_{vv}(u,v,w)\ dv^2+g_{ww}(u,v,w)\ dw^2\nonumber\\&\ &+2g_{uv}(u,v,w)\ dudv+2g_{uw}(u,v,w)\ dudw+2g_{vw}(u,v,w)\ dvdw,
\end{eqnarray}
which represents a static (3+1)-d curved traversable wormhole, with $t$ denoting the time coordinate. Just like the lower-dimensional version, this method assures that the metric coefficients are all smooth functions of the spatial coordinates, with no singularities.

It has already been proven in \cite{Vee2012} that the 3-manifold of revolution around a straight line yields the spherically symmetric solution, which was the focus of Morris and Thorne's original study \cite{Kip}. Also shown was that the spacetime metric of the (3+1)-d helical wormhole has two spatial cross-terms compared to one in its lower-dimensional counterpart, where cross-terms are notorious for making the calculation of the field tensors tedious. The (2+1)-d catenary wormhole has no cross-term since it is a plane curve, with a proof that this is true for any surface of revolution around plane curves (where the orthonormal vectors are mutually perpendicular to the tangent vector) given in \cite{Vee2012}. The same holds for the (3+1)-d version, as is shown in the appendix.

The goal of constructing curved traversable wormholes is to carefully distribute the mass-energy so that there exists a connected region through the wormhole supported locally by ordinary matter. This would then allow humans to traverse safely without direct contact with the exotic matter. The physical interpretations from the (2+1)-d study concluded the existence of safe geodesics through properly engineered curved traversable wormholes. Repeating the full calculations in (3+1)-d would come with much greater complications, as already noted by Morris and Thorne \cite{Kip} which is why they only focused on the spherically symmetric solutions. To minimise the algebraic distractions for the (3+1)-d version to show that indeed a properly built curved traversable wormhole in this universe admits safe geodesics, we shall assume zero tidal force right from the start so that $\Phi(u,v,w)=0$.

We have two examples of (3+1)-d curved traversable wormholes, viz. the catenary wormhole obtained from a plane curve (the derivation of the metric is given in Section 4), and the helical wormhole \cite{Vee2012} obtained from a space or 3-d curve (in fact in (3+1)-d spacetime, it is possible to construct one from a 4-d hyperspace curve, if desired). As our aim here is to show the possibility of curved traversable wormholes possessing safe geodesics, we are going to focus only on the catenary wormhole since it is free from cross-terms.

\section{The physics of the zero tidal force (3+1)-d wormholes (without cross-terms) according to the Einstein field equations, and comparison with (2+1)-d}
\label{sec3}

The physics that governs the curved traversable wormholes is Einstein's theory of general relativity with the field equations
\begin{eqnarray}
G_{\mu\nu}=R_{\mu\nu}-\frac{1}{2}g_{\mu\nu}R=8\pi T_{\mu\nu}.
\end{eqnarray}
Greek indices run from 0, 1, 2 and 3, Einstein summation convention is assumed, and geometrised units shall be adopted. For the spacetime metric of the form
\begin{eqnarray}\label{met1}
ds^2=-dt^2+g_{11}(u,v,w)du^2+g_{22}(u,v,w)dv^2+g_{33}(u,v,w)dw^2,
\end{eqnarray}
the Christoffel symbols, Ricci curvature tensor and Ricci scalar can be calculated. For this metric, the non-zero components of the Ricci curvature tensor $R_{\mu\nu}$ are those where both $\mu,\nu\in\{1,2,3\}$. The Einstein tensor in the proper reference frame where the observer is always at rest is $G_{\hat{\mu}\hat{\nu}}=R_{\hat{\mu}\hat{\nu}}-\frac{1}{2}g_{\hat{\mu}\hat{\nu}}R$ (with $g_{\hat{\mu}\hat{\nu}}$ being Minkowskian). The components of $G_{\hat{\mu}\hat{\nu}}$ in terms of the given metric tensor (eq. (\ref{met1})) can be shown to conform to the following rules:
\begin{enumerate}
\item $G_{\hat{\mu}\hat{0}}=G_{\hat{0}\hat{\mu}}=\frac{1}{2}\delta_{\hat{\mu}\hat{0}}R$, where $\delta_{\hat{\mu}\hat{\nu}}$ is the Kronecker delta.
\item For the spatial diagonal terms,
\begin{eqnarray}\label{SET1}
G_{\hat{1}\hat{1}}&=&\frac{g_{22,33}+g_{33,22}}{2g_{22}g_{33}}+\frac{g_{22,1}g_{33,1}}{4g_{11}g_{22}g_{33}}\nonumber\\&\ &-\frac{1}{4g_{22}g_{33}}\left[\frac{g_{22,2}g_{33,2}+(g_{22,3})^2}{g_{22}}+\frac{g_{22,3}g_{33,3}+(g_{33,2})^2}{g_{33}}\right],
\end{eqnarray}
and $G_{\hat{2}\hat{2}}$ is obtained from $G_{\hat{1}\hat{1}}$ by replacing $\hat{1}\rightarrow\hat{2}$, $\hat{2}\rightarrow\hat{3}$ and $\hat{3}\rightarrow\hat{1}$ respectively, with $G_{\hat{3}\hat{3}}$ obtained from $G_{\hat{2}\hat{2}}$ in a similar manner.
\item For the spatial non-diagonal terms,
\begin{eqnarray}\label{SET2}
G_{\hat{1}\hat{2}}&=&G_{\hat{2}\hat{1}}=\frac{1}{4g_{33}\sqrt{g_{11}g_{22}}}\left(\frac{g_{11,2}g_{33,1}}{g_{11}}+\frac{g_{22,1}g_{33,2}}{g_{22}}+\frac{g_{33,1}g_{33,2}}{g_{33}}\right)\nonumber\\&\ &-\frac{g_{33,12}}{2g_{33}\sqrt{g_{11}g_{22}}},
\end{eqnarray}
and $G_{\hat{2}\hat{3}}=G_{\hat{3}\hat{2}}$ is obtained from $G_{\hat{1}\hat{2}}=G_{\hat{2}\hat{1}}$ by replacing $\hat{1}\rightarrow\hat{2}$, $\hat{2}\rightarrow\hat{3}$ and $\hat{3}\rightarrow\hat{1}$ respectively, with $G_{\hat{3}\hat{1}}=G_{\hat{1}\hat{3}}$ obtained from $G_{\hat{2}\hat{3}}=G_{\hat{3}\hat{2}}$ through a similar permutation of indices.
\end{enumerate}
Note also that
\begin{eqnarray}
G_{\hat{1}\hat{1}}+G_{\hat{2}\hat{2}}+G_{\hat{3}\hat{3}}&=&\left(R_{\hat{1}\hat{1}}-\frac{1}{2}R\right)+\left(R_{\hat{2}\hat{2}}-\frac{1}{2}R\right)+\left(R_{\hat{3}\hat{3}}-\frac{1}{2}R\right)\\
&=&(R_{\hat{1}\hat{1}}+R_{\hat{2}\hat{2}}+R_{\hat{3}\hat{3}})-\frac{3}{2}R\\
&=&R-\frac{3}{2}R\\
&=&-G_{\hat{0}\hat{0}},
\end{eqnarray}
where we have used $R=g^{\hat{\mu}\hat{\nu}}R_{\hat{\mu}\hat{\nu}}=R_{\hat{1}\hat{1}}+R_{\hat{2}\hat{2}}+R_{\hat{3}\hat{3}}$. In other words, $\displaystyle\sum_{\hat{\mu}=\hat{0}}^{\hat{3}}{G_{\hat{\mu}\hat{\mu}}}=0$. So essentially, $G_{\hat{0}\hat{0}}$ can be obtained from just remembering the spatial diagonal terms. Also, this implies that if the non-diagonal terms are all zero and the null energy condition is satisfied, then the strong energy condition is automatically satisfied (see pages 115-116 of \cite{Vis2} for the definitions of the energy conditions). It is straightforward to check the spherically symmetric case in \cite{Kip} that in the absence of the tidal force $\Phi$ as it is assumed here, the sum of the diagonal terms of the Einstein tensor in the proper reference frame is zero.

Comparisons can be made with the (2+1)-d version which is discussed in \cite{Vee2012}. In that lower-dimensional analysis for the general metric including spatial cross-terms, all terms in the Einstein tensor vanish in the absence of tidal force except for the $G_{00}=\frac{1}{2}R$ term. Here, the (3+1)-d version for the metric without cross-terms already consists of non-zero spatial diagonal and non-diagonal terms for $G_{\hat{\mu}\hat{\nu}}$. Close inspection of the terms reveals that these result from the third spatial coordinate which does not exist in a (2+1)-d universe, so that $g_{33}$ would be zero and the partial derivative of any component of the metric tensor with respect to the third spatial coordinate is zero. Thus the vanishing of all forms of radial tension and lateral pressure in the lower-dimensional spacetime turns out to be a special case where there are no such terms involving the mixing of all three spatial coordinates. Note also that the lower-dimensional spacetime does not satisfy the property $\displaystyle\sum_{\mu=0}^{2}{G_{\mu\mu}}=0$ since $G_{11}$ and $G_{22}$ are identically zero for any point in spacetime, whereas the Ricci scalar is not always zero for a curved spacetime.

The (3+1)-d spacetime is the lowest dimension where these ``mixing terms'' appear which yield a much richer physics. Instead of dust which exerts no pressure on neighbouring elements for the lower-dimensional case, a (3+1)-d curved wormhole has to be supported by a general fluid that exerts pressures and shear forces according to Eqs. (\ref{SET1}-\ref{SET2}). It would be interesting to see even more complicated mixing of the spatial coordinates in higher-dimensional spacetimes, leading to perhaps new physics that is absent in our (3+1)-d universe.

\section{The catenary wormhole}
\label{sec4}

Consider the plane curve $\vec{\psi}_c(v)=(v,\cosh{v},0,0)$. The unit tangent vector is $(\sech{v},\tanh{v},0,0)$. Following the construction done in \cite{Vee2012} for the (2+1)-d catenary wormhole with the same radial function $Z(v)=\frac{v^4+v^2+1}{8(v^2+1)}$, the three orthonormal vectors shall be chosen as $\vec{n}_1(v)=(-\tanh{v},\sech{v},0,0)$, $\vec{n}_2(v)=(0,0,1,0)$ and $\vec{n}_3(v)=(0,0,0,1)$. The resulting spacetime metric (recall the zero tidal force condition imposed at the end of Section 2) would be
\begin{eqnarray}\label{cat}
ds^2=-dt^2+g_{vv}(u,v)\ dv^2+Z(v)^2(du^2+\sin^2{u}\ dw^2),
\end{eqnarray}
where $g_{vv}(u,v)$ is the same as Eq. (49) in \cite{Vee2012}.

It is worth mentioning that the metric of a 3-manifold of revolution around a plane curve takes the general form of Eq. (\ref{cat}), where $g_{vv}$ is some function of two spatial variables $u$ and $v$. Taking a slice where the third variable $w$ is a constant reduces to the metric for the corresponding surface of revolution around the same plane curve. This is similar to the (3+1)-d spherically symmetric case \cite{Kip} where any slice with constant $\phi$ (the azimuthal angle) reduces to the same (2+1)-d metric \cite{2tw1}. The difference between using a straight line and a plane curve is that the $g_{vv}$ term for the former depends only on $v$ (the parameter for points along the given straight line) due to the spherical symmetry, in constrast to the dependence on $u$ and $v$ for the latter (the parameters for the ``loop'' around the given curve as well as for the points along the given curve) due to the lack of spherical symmetry. Moreover for space curves, the metric for the (3+1)-d helical wormhole with the two non-zero spatial cross-terms \cite{Vee2012} indicates even further deviation from the spherical symmetry.

Using the results of the previous section, the non-zero components of the Einstein tensor in the proper reference frame ($\{0,1,2,3\}$ corresponds to $\{t,u,v,w\}$) for the catenary wormhole with metric in Eq. (\ref{cat}) are
\begin{eqnarray}\label{SETcat1}
G_{\hat{0}\hat{0}}&=&-\frac{(g_{11,22}+g_{22,11})}{2g_{11}g_{22}}-\frac{g_{33,11}}{2g_{11}g_{33}}-\frac{g_{33,22}}{2g_{22}g_{33}}\nonumber\\&\ &+\frac{1}{4g_{11}g_{22}}\left[\frac{(g_{11,2})^2}{g_{11}}+\frac{g_{11,2}g_{22,2}+(g_{22,1})^2}{g_{22}}\right]+\frac{(g_{33,1})^2}{4g_{11}(g_{33})^2}\nonumber\\&\ &+\frac{1}{4g_{22}g_{33}}\left[\frac{g_{22,2}g_{33,2}}{g_{22}}+\frac{(g_{33,2})^2}{g_{33}}\right]-\frac{(g_{11,2}g_{33,2}+g_{22,1}g_{33,1})}{4g_{11}g_{22}g_{33}}\\
G_{\hat{1}\hat{1}}&=&\frac{g_{33,22}}{2g_{22}g_{33}}-\frac{1}{4g_{22}g_{33}}\left[\frac{g_{22,2}g_{33,2}}{g_{22}}+\frac{(g_{33,2})^2}{g_{33}}\right]+\frac{g_{22,1}g_{33,1}}{4g_{11}g_{22}g_{33}}\\
G_{\hat{2}\hat{2}}&=&\frac{g_{33,11}}{2g_{11}g_{33}}-\frac{(g_{33,1})^2}{4g_{11}(g_{33})^2}+\frac{g_{11,2}g_{33,2}}{4g_{11}g_{22}g_{33}}\\
G_{\hat{3}\hat{3}}&=&\frac{g_{11,22}+g_{22,11}}{2g_{11}g_{22}}-\frac{1}{4g_{11}g_{22}}\left[\frac{(g_{11,2})^2}{g_{11}}+\frac{g_{11,2}g_{22,2}+(g_{22,1})^2}{g_{22}}\right]\\\label{SETcat2}
G_{\hat{1}\hat{2}}&=&G_{\hat{2}\hat{1}}=\frac{1}{4g_{33}\sqrt{g_{11}g_{22}}}\left(\frac{g_{11,2}g_{33,1}}{g_{11}}+\frac{g_{22,1}g_{33,2}}{g_{22}}+\frac{g_{33,1}g_{33,2}}{g_{33}}\right)\nonumber\\&\ &-\frac{g_{33,12}}{2g_{33}\sqrt{g_{11}g_{22}}}.
\end{eqnarray}
With these, the components of the field equations are plotted as functions of $u$ and $v$ in Figs. 1-5, as they are all independent of $w$. For any particular $v$, the variation of the components of the field tensor over $u$ can be obtained, and this would be the same for all $w\in[0,2\pi]$.

By construction, $w$ is the azimuthal coordinate that runs from $0$ to $2\pi$, and so the catenary wormhole has azimuthal symmetry. The $u$-coordinate is the polar coordinate that runs from $0$ to $\pi$, whilst $v\in\R$ describes the points along the catenary curve. Similar to the lower-dimensional case, $u=0$ refers to the innermost part of the catenary wormhole, whilst $u=\pi$ refers to the outermost part. This can be checked explicitly by substituting $u=0$ or $u=\pi$ into Eq. (\ref{eq1}) for the 3-manifold of revolution around the catenary curve (as defined in the paragraph containing Eq. (\ref{cat})) and inspecting the geometry (Fig. 6 in \cite{Vee2012} may aid in visualising the (3+1)-d version). At either of these two points, the $w$-coordinate becomes degenerate such that all $w\in[0,2\pi]$ correspond to each of the two poles of the 2-sphere. The ``throat'' is the part of the wormhole with the smallest radius, corresponding to $v=0$. Unlike the spherically symmetric case where the throat is completely supported by exotic matter \cite{Kip}, the throat of the catenary wormhole is merely the location where it has the smallest radius. The construction of a curved traversable wormhole that breaks the spherical symmetry allows for a continuous region through the wormhole that is supported locally by non-exotic material, as we will show below.

\subsection{Properties of the mass-energy fluid required to support the catenary wormhole}

The construction of the spacetime geometry of the catenary wormhole determines the stress-energy tensor via the field equations. Therefore, builders of the catenary wormhole would have to search or synthesise the appropriate materials that have the properties as demanded by Eqs. (\ref{SETcat1}-\ref{SETcat2}). As already noted in the previous section, a general fluid that exerts pressures and shear stress is required, in contrast to dust (for the lower-dimensional spacetime). The term $G_{\hat{1}\hat{2}}=G_{\hat{2}\hat{1}}$ represents the fact that the fluid elements exert a shear force (this is the only shear force since other non-diagonal $G_{\hat{i}\hat{j}}$ terms are zero, where $\hat{i},\hat{j}\in\{\hat{1},\hat{2},\hat{3}\}$). Nevertheless as can be observed in Fig. 5, such fluid elements occur in very localised regions near the throat, as such shear force dies off very rapidly as one goes away along the $v$-direction. Furthermore, the maximum strength is relatively small compared to that of the principal pressures that they exert, viz. about 28 times smaller than the maximum magnitude of the pressures along $u$ and $w$ (the ``angular directions''), as well as being about 191 times smaller than that of the pressure along $v$ (the direction along the catenary curve). Note also that the fluid elements do not exert any shear force for $u=0$ and $u=\pi$. The shear force is also absent at the throat $v=0$. Although this shear force may seem feeble compared to the principal pressures, the very fact that it is non-zero represents one key feature of a (3+1)-d non-spherically symmetric geometry, as it is completely absent in spherically symmetric solutions \cite{Kip}.

The strong principal pressures that these fluid elements exert on their neighbours are also localised near the throat of the catenary wormhole where the Ricci scalar of spacetime is the greatest, and decay very quickly away from it along the $v$-direction. (See Figs. 1-4.) The pressures along the two angular directions $u$ and $w$ are highly similar where they vary from positive at $u=0$ indicating that the fluid elements are under compression, to negative at $u=\pi$ so that these fluid elements are under tension, though their expressions $G_{\hat{1}\hat{1}}$ and $G_{\hat{3}\hat{3}}$ are not identical. This illustrates a second key feature of a non-spherically symmetric geometry. Spherically symmetric wormholes have equivalent angular pressures \cite{Kip}, since the spherical symmetry of a spacetime would imply that there is no preferred angular direction and hence the physics along the two angular coordinates must be the same. Furthermore, the terms in the stress-energy tensor for the catenary wormhole generally depend on $u$ and $v$, unlike the spherically symmetric ones that only depend on the radial coordinate $r$ which defines the shape function (just like $v$ in our case). Anyway, the pressure exerted along the $v$-direction is always negative such that the fluid elements are under tension along the direction of the catenary curve, with a maximum strength at the throat that is about 7 times greater than that of the other two principal pressures. Fig. 6 shows the forces experienced by fluid elements located at the throat of the catenary wormhole where the shear stress is absent. The $w$-coordinate is suppresed, taking advantage of the azimuthal symmetry to aid in visualisation. The corresponding picture for a spherically symmetric one is also shown for comparison in Fig. 7.

The lower-dimensional catenary wormhole is supported by dust of positive and negative mass-energy density \cite{Vee2012}. The presence of negative mass-energy dust is of no surprise because it is the consequence of a traversable wormhole having to violate the null energy condition \cite{Vis2,Vis4,TCT,Vis3,Vis5}. Since there is no pressure and no shear forces at all, violation of the null energy condition takes place only when the mass-energy dust is negative. It turns out that this is not true for the (3+1)-d version as is clearly depicted in Fig. 1: the fluid elements are all of positive mass-energy density. The outer part of the (3+1)-d catenary wormhole (where $u=\pi$) is denser, corresponding to the region with the greatest Ricci scalar. The inner part (where $u=0$) on the other hand also has positive Ricci scalar, but is not as dense as the outer part. In contrast, the inner part of the (2+1)-d version has negative Ricci scalar and that is the region supported by negative mass-energy dust that violates the null energy condition. Violation of the null energy condition in the inner part of the (3+1)-d catenary wormhole is evident in Fig. 9 where $G_{\hat{0}\hat{0}}+G_{\hat{2}\hat{2}}<0$ around that area.

\subsection{A safe geodesic through the catenary wormhole}

A safe geodesic for the lower-dimensional catenary wormhole was shown to be the one where $u=\pi$, corresponding to the outermost extreme of the wormhole \cite{Vee2012}. The same turns out to be true for the (3+1)-d counterpart, as we will now show.

The non-zero Christoffel symbols for the catenary wormhole are $\Gamma^1_{22}(u,v)$, $\Gamma^1_{33}(u)$, $\Gamma^1_{12}(v)=\Gamma^1_{21}(v)$, $\Gamma^2_{11}(u,v)$, $\Gamma^2_{22}(u,v)$, $\Gamma^2_{33}(u,v)$, $\Gamma^2_{12}(u,v)=\Gamma^2_{21}(u,v)$, $\Gamma^3_{13}(u)=\Gamma^3_{31}(u$), $\Gamma^3_{23}(v)=\Gamma^3_{32}(v)$, where their explicit dependence on the spatial variables are shown. The geodesic equations are
\begin{eqnarray}
\ddot{u}+\Gamma^1_{11}\dot{u}^2+\Gamma^1_{22}\dot{v}^2+\Gamma^1_{33}\dot{w}^2+2\Gamma^1_{12}\dot{u}\dot{v}+2\Gamma^1_{13}\dot{u}\dot{w}+2\Gamma^1_{23}\dot{v}\dot{w}=0\\
\ddot{v}+\Gamma^2_{11}\dot{u}^2+\Gamma^2_{22}\dot{v}^2+\Gamma^2_{33}\dot{w}^2+2\Gamma^2_{12}\dot{u}\dot{v}+2\Gamma^2_{13}\dot{u}\dot{w}+2\Gamma^2_{23}\dot{v}\dot{w}=0\\
\ddot{w}+\Gamma^3_{11}\dot{u}^2+\Gamma^3_{22}\dot{v}^2+\Gamma^3_{33}\dot{w}^2+2\Gamma^3_{12}\dot{u}\dot{v}+2\Gamma^3_{13}\dot{u}\dot{w}+2\Gamma^3_{23}\dot{v}\dot{w}=0,
\end{eqnarray}
where $\psi(\zeta)=\sigma(u(\zeta),v(\zeta),w(\zeta))$ is a geodesic on the manifold if it satisfies the geodesic equations.

Consider the initial condition $u_0=\pi$, $v_0\in\R$, $w_0\in[0,2\pi]$, $\dot{u}_0=0$, $\dot{v}_0\in\R\backslash\{0\}$, $\dot{w}_0=0$. Note that $|\dot{v}_0|<1$ corresponds to a massive object moving slower than the speed of light (recall that we are using geometrised units), $|\dot{v}_0|=1$ represents a massless particle going at the speed of light, and $|\dot{v}_0|>1$ would be for tachyons (if such particles exist) travelling faster than the speed of light. When $u_0=\pi$, the Christoffel symbols $\Gamma^1_{22}(\pi,v)$, $\Gamma^1_{33}(\pi)$, $\Gamma^2_{33}(\pi,v)$, $\Gamma^2_{12}(\pi,v)=\Gamma^2_{21}(\pi,v)$ become zero. The geodesic equations then give
\begin{eqnarray}
\ddot{u}_0&=&0\\
\ddot{v}_0&=&-[\Gamma^2_{22}(\pi,v_0)](\dot{v}_0)^2\\
\ddot{w}_0&=&0.
\end{eqnarray}
For an infinitesimal increase $d\zeta$,
\begin{eqnarray}
u&=&u_0+\dot{u}_0d\zeta=\pi\\
\dot{u}&=&\dot{u}_0+\ddot{u}_0d\zeta=0\\
v&=&v_0+\dot{v}_0d\zeta\\
\dot{v}&=&\dot{v}_0+\ddot{v}_0d\zeta=\dot{v}_0-[\Gamma^2_{22}(\pi,v_0)](\dot{v}_0)^2d\zeta\\
w&=&w_0+\dot{w}_0d\zeta=w_0\\
\dot{w}&=&\dot{w}_0+\ddot{w}_0d\zeta=0.
\end{eqnarray}
This implies that $\psi(\zeta)=\sigma(\pi,v(\zeta),w_0)$ is a geodesic on the catenary wormhole, where $w_0\in[0,2\pi]$ is a constant. The point where $u=\pi$ corresponds to $w$ becoming degenerate, such that all $w_0\in[0,2\pi]$ refer to the same extremal point.

To verify that this is a safe geodesic, we need to check that it satisfies the null energy condition \cite{Vis2}. As noted, the shear stress $G_{\hat{1}\hat{2}}=G_{\hat{2}\hat{1}}$ vanishes at $u=\pi$, so that $G_{\hat{\mu}\hat{\nu}}$ is diagonal. The null energy condition requires that $G_{\hat{0}\hat{0}}+G_{\hat{1}\hat{1}}\geq0$, $G_{\hat{0}\hat{0}}+G_{\hat{2}\hat{2}}\geq0$ and $G_{\hat{0}\hat{0}}+G_{\hat{3}\hat{3}}\geq0$. The relevant graphs are plotted as functions of $u$ and $v$ in Figs. 8-10, as well as the variations with respect to $v$ at $u=\pi$ in Figs. 11-13. The result is affirmative that this geodesic indeed satisfies the null energy condition, similar to the lower-dimensional case.

(Note: There is a minor error in our (2+1)-d paper \cite{Vee2012} on showing that the outermost curve through the (2+1)-d catenary wormhole is a geodesic, where we mentioned that all the relevant Christoffel symbols vanish at $u=0$ and $u=\pi$. This is a slight oversight in our calculations, where not all of them vanish. Nevertheless upon rectifying the calculation error, it turns out that $\vec{c}_\pi$ is indeed a geodesic by means of plotting it out numerically. This can also be shown by considering the local change in the position and direction of a point $u_0=\pi$, $v_0\in\R$, with initial velocity $\dot{u}_0=0$, $\dot{v}_0\neq0$. The conclusion agrees with the geometrical argument that we gave in that paper.)

\subsection{The radial function}

Just like the (2+1)-d version, the radial function is designed such that the Ricci scalar, all terms in the Ricci curvature tensor and the Riemann curvature tensor go to zero at infinity so that the spacetime grows asymptotically flat away from the wormhole. Our choice of the radial function $Z(v)=\frac{v^4+v^2+1}{8(v^2+1)}$ for the catenary wormhole (both (2+1)-d and (3+1)-d) may appear to be somewhat arbitrary. There can certainly be other possibilities of the radial function, and that we define is one that permits the existence of a continuous region through the wormhole supported locally by ordinary material.

The flaring out of the wormhole where it grows asymptotically flat at vast distances away from the throat would contribute towards negative curvature such that exotic matter is required to support such a region. This is especially conspicuous in the (2+1)-d version where the field equations reduce to only one equation: mass-energy density is proportional to Ricci scalar. The key idea of prescribing a useful radial function would be to produce cancellation of the negative contribution due to flaring out by the positive contribution due to the curvature of the catenary curve itself. This has successfully allowed us to constrain the exotic matter to within a bounded region along the $u$- and $w$-directions, opening up that continuous region through the catenary wormhole that is free from it.

It would be interesting and useful to try to further constrain the use of exotic matter along the $v$-direction so that the need for exotic matter would truly be of finite quantity, since synthesis of such material is quite difficult (at least for the time being). A careful observation would show that a faster rate of flaring out would demand greater use of such material to promote more negative contribution to the curvature, whilst using less exotic material would slow down the rate of flaring out.

In fact, the required mass-energy can be truncated and confined to within a bounded region containing the wormhole's throat, something which has also been discussed for the lower-dimensional wormholes \cite{Vee2012}. Essentially, a region of mass-energy with the prescribed distribution for a region containing the throat of the catenary wormhole can be constructed, with the regions outside being vacuum. This is akin to a massive interstellar body that generates spacetime curvature, and the vacuum regions outside are Ricci flat but not Riemann flat (though the Riemann curvature gets asymptotically flat). Unlike the usual interstellar objects out in the sky that we have discovered, this one must be partly exotic and partly ordinary. Travellers would then go through the ordinary part of this body to traverse the wormhole. As was mentioned in Morris and Thorne's original work \cite{Kip}, ``Traveller must not couple strongly to material that generates wormhole curvature (wormhole must be threaded by a vacuum tube through which she moves, or wormhole material must be of type that couples weakly to ordinary matter).''

In order to find the radial function of the spacetime for the vacuum regions outside the catenary wormhole, one would have to solve the vacuum field equations and match it to the boundary condition where the mass-energy is truncated. This is easily done in (2+1)-d \cite{Vee2012}, since vacuum in (2+1)-d spacetime necessarily demands that the spacetime is Riemann flat. The exterior regions are hence truncated cones that are matched at the boundaries of the catenary wormhole where the mass-energy is cut off.

This is not so trivial in (3+1)-d spacetime, since being Ricci flat (vacuum) does not imply that the full Riemann curvature is zero. Therefore the exterior vacuum regions are not flat, just like the Riemann curvature of the exterior solution of a spherically symmetric star being non-zero. Things are even more complicated for (3+1)-d spacetime because unlike the Schwarzschild solution with spherical symmetry, there is no a priori reason to expect that the vacuum part of the spacetime would take the form of Eq. (\ref{eq1}) where it is a manifold of revolution around the catenary curve. Even if we assume this, the problem then evolves into the arduous task of solving the vacuum field equations given by Eqs. (\ref{SETcat1}-\ref{SETcat2}) where $G_{\hat{\mu}\hat{\nu}}=0$. It does not seem that an analytic solution to these equations can be found easily, especially with the requirement of having to satisfy the boundary condition at the part where the wormhole material is truncated. Moreover if the vacuum region outside the catenary wormhole is not a manifold of revolution around the catenary curve, one would then face the onerous task of solving the full vacuum field equations with the required boundary condition. One may turn to numerically solving the radial function (if it takes the form of a manifold of revolution) or the full Einstein field equations for the vacuum region, however this would defeat the whole purpose of our method of constructing the spacetime geometry, versus having to solve those partial differential equations.

\section{Advantages and potential of this geometrical method in visualising the spacetime geometry and finding new solutions to the field equations of general relativity}
\label{sec5}

The technique in building static curved traversable wormholes is based upon the specification of the geometry of spacetime and then using the Einstein field equations to determine the physics of the wormhole, viz. the mass-energy density, the pressures and stresses to support it. More precisely, we have elaborated on how to generate manifolds of revolution around a given smooth curve. By relaxing the constraints on the usual physical materials and allowing the actual construction of the wormhole to be supported by matter of the desired property (this is the philosophy adopted by Morris and Thorne's original study \cite{Kip}), we are free to design spacetimes with properties like being singularity-free and having no tidal force.

The Einstein field equations are a set of highly nonlinear partial differential equations. Solving them analytically is no easy task, with numerical relativity having its own challenges (see for instance \cite{num1} and \cite{num2}). The construction of spacetime on the other hand requires no differential equations to be solved. Instead, it needs only taking partial derivatives and carrying out elementary algebra. This may seem to be too much freedom that would lead to non-physical solutions, nevertheless whether a solution is physical or not ultimately boils down to experimental observations. Quantum field theory allows the existence of exotic matter \cite{Exo1,Exo2,Exo3}, hence traversable wormholes are not unphysical. While it may be technologically impossible for us to build a wormhole today, it may turn out to be only a matter of time before intergalactic travel becomes the norm in human civilisation.

In fact, one may attempt the construction of a dynamic spacetime by allowing the geometry to progress with time. As a simple example to illustrate this point, consider the 3-manifold of revolution generated around the time-dependent straight line $\vec{\psi}_l(t,v)=(0,0,0,\Omega(t)z(v))$. The $\Omega(t)$ factor implies that the distance between any two points on the line will evolve with time as prescribed by that factor. We shall take the three orthonormal vectors mutually perpendicular to $\vec{\psi}_l$ as the three coordinate axes $\vec{e}_1$, $\vec{e}_2$ and $\vec{e}_3$, with the time-dependent radial function $\Omega(t)Z(v)$. (Both $\Omega(t)>0$ and $Z(v)>0$.) The resulting 3-manifold that is generated would be
\begin{eqnarray}
\vec{\sigma}_l(t,u,v,w)&=&(\Omega(t)Z(v)\cos{u},\Omega(t)Z(v)\sin{u}\cos{w},\Omega(t)Z(v)\sin{u}\sin{w},\nonumber\\&\ &\Omega(t)z(v)).
\end{eqnarray}
The geometrical picture of this construction would be a spherically symmetric wormhole whose overall geometry expands or contracts, according to $\Omega(t)$. Direct computation of the spatial metric leads to
\begin{eqnarray}
ds^2=\Omega(t)^2\left[\left(Z'(v)^2+z'(v)^2\right)dv^2+Z(v)^2(du^2+\sin{u}^2\ dw^2)\right],
\end{eqnarray}
so that the spacetime metric would take the form
\begin{eqnarray}
ds^2&=&-e^{2\Phi}c^2dt^2+\Omega(t)^2[\left(Z'(v)^2+z'(v)^2\right)dv^2\nonumber\\&\ &+Z(v)^2(du^2+\sin{u}^2\ dw^2)],
\end{eqnarray}
where $\Phi$ may be prescribed to be a function of just the $v$-coordinate. This can then be adapted to represent inflating and evolving wormholes as studied in \cite{inf,inf2,inf3}.

The key advantage of this method is that it provides the clear picture of the spacetime that is being constructed since the method itself exactly describes the geometry of spacetime thus overcoming our limitation in trying to visualise higher-dimensional manifolds, as opposed to solving the field equations and resorting to taking slices of the metric. Dynamical spacetimes are typically much more difficult to deal with as compared to static ones. We shall not delve further here since the focal point of this paper is on constructing static curved traversable wormholes that allow safe geodesics through them. A thorough discussion on more general time-dependent wormholes would be the subject of future research.

Apart from that, this approach of building the geometry of spacetime can be applied to generalise existing solutions in general relativity. For instance, one may possibly analyse the Schwarzschild solution by perturbing the geometry of spacetime instead of perturbing the terms in the stress-energy tensor. One can try to manipulate the curvature of the Schwarzschild spacetime by constructing it using this method and study the effects on the spherically symmetric mass that gives rise to such a spacetime, via the field equations. We hope to report on the details in a future study.

Recently, the quantum gravitational phenomenon of a firewall near the horizon of a black hole has been a hotly debated topic \cite{nature,Firewall1}. One intriguing proposal related to this issue has to do with the idea that when two particles are entangled, they are actually connected by a tiny wormhole \cite{Susskindwormhole}. It would be interesting to see how this will further develop, in an attempt to resolve the paradox on black hole and information \cite{BHW,Haw1,Susskinda,Hooft,Susskindb}, and how wormholes may eventually play a pivotal role.

\section{Concluding remarks}
\label{sec6}

We have described the method of constructing curved traversable wormholes in (3+1)-d spacetime, building on the previous lower-dimensional study in \cite{Vee2012}. The static zero tidal force (3+1)-d catenary wormhole shows that one can indeed design curved traversable wormholes that admit safe geodesics through them supported locally by ordinary matter, representing a new class of such wormholes in addition to Visser's polyhedral wormhole \cite{vis1} and Teo's rotating wormhole \cite{teo}. This is thus consistent with the lower-dimensional study, so that (3+1)-d curved traversable wormholes allow humans or some advanced (3+1)-d civilisations to perhaps some day be able to traverse them without direct contact with exotic matter. The calculations of the field equations also provided insights between the physics of (2+1)-d general relativity versus (3+1)-d general relativity, with the added spatial dimension yielding much richer physics due to a more general fluid instead of dust in the lower dimension. Moreover, this method of constructing the spacetime first and then applying the field equations to uncover the physics would provide an important avenue in searching for undiscovered solutions and extending known ones in general relativity. Indeed the freedom offered by this approach may generate unphysical solutions, nevertheless it could serve as an important tool in understanding the geometry of spacetime.

\begin{acknowledgements}
We would like to extend our deepest gratitute to Meng Lee Leek from Nanyang Technological University for the highly valuable discussion as well as his time and effort in carefully reviewing our draft. We wish to also express our appreciation to Sampsa Vihonen from University of Jyv\"{a}skyl\"{a} for his useful comments. Furthermore, special thanks is directed to the reviewers whose comments help us improve the paper.
\end{acknowledgements}

\section{Appendix: Proof of metric for 3-manifold of revolution generated around plane curves being free of spatial cross-terms}

Consider a unit-speed plane curve in $\R^4$, $\vec{\psi}(v)=(f(v),g(v),0,0)$, where $f'(v)$ and $g'(v)$ are not simultaneously zero. A unit-speed curve has unit tangent vector, i.e. $f'(v)^2+g'(v)^2=1$ for all $v\in\R$. Similar to the lower-dimensional proof \cite{Vee2012}, it is enough to consider unit-speed curves since any smooth or regular curve can be reparametrised to be unit-speed.

With the three orthonormal vectors (all perpendicular to the unit tangent vector) being chosen as $\vec{n}_1(v)=(-g'(v),f'(v),0,0)$, $\vec{n}_2(v)=(0,0,1,0)$ and $\vec{n}_3(v)=(0,0,0,1)$, the 3-manifold of revolution around $\vec{\psi}(v)$ is
\begin{eqnarray}
\vec{\sigma}(u,v,w)=
\left(
\begin{array}{c}
f(v)-g'(v)Z(v)\cos{u}\\
g(v)+f'(v)Z(v)\cos{u}\\
Z(v)\sin{u}\cos{w}\\
Z(v)\sin{u}\sin{w}
\end{array}
\right).
\end{eqnarray}
The partial derivatives of $\vec{\sigma}$ are then
\begin{eqnarray}
\vec{\sigma}_u(u,v,w)=
\left(
\begin{array}{c}
g'(v)Z(v)\sin{u}\\
-f'(v)Z(v)\sin{u}\\
Z(v)\cos{u}\cos{w}\\
Z(v)\cos{u}\sin{w}
\end{array}
\right),
\end{eqnarray}
\begin{eqnarray}
\vec{\sigma}_v(u,v,w)=
\left(
\begin{array}{c}
f'(v)-g'(v)Z'(v)\cos{u}-g''(v)Z(v)\cos{u}\\
g'(v)+f'(v)Z'(v)\cos{u}+f''(v)Z(v)\cos{u}\\
Z'(v)\sin{u}\cos{w}\\
Z'(v)\sin{u}\sin{w}
\end{array}
\right),
\end{eqnarray}
\begin{eqnarray}
\vec{\sigma}_w(u,v,w)=
\left(
\begin{array}{c}
0\\
0\\
-Z(v)\sin{u}\sin{w}\\
Z(v)\sin{u}\cos{w}
\end{array}
\right).
\end{eqnarray}
With this, it is easy to see that $g_{uw}=\vec{\sigma}_u\cdot\vec{\sigma}_w$ and $g_{vw}=\vec{\sigma}_v\cdot\vec{\sigma}_w$ are zero. Also, $g_{uv}=\vec{\sigma}_u\cdot\vec{\sigma}_v$ turns out to be the same as Eq. (58) in \cite{Vee2012} which is zero due to the unit-speed condition. This completes the proof.

Furthermore, $g_{uu}=Z(v)^2$ and $g_{vv}(u,v)$ are identical to that of the lower-dimensional counterpart, with $g_{ww}=Z(v)^2\sin^2{u}$.

\bibliographystyle{spphys}       
\bibliography{Citation}   

\newpage
\begin{figure}[h]
\centering
\includegraphics[width=10cm]{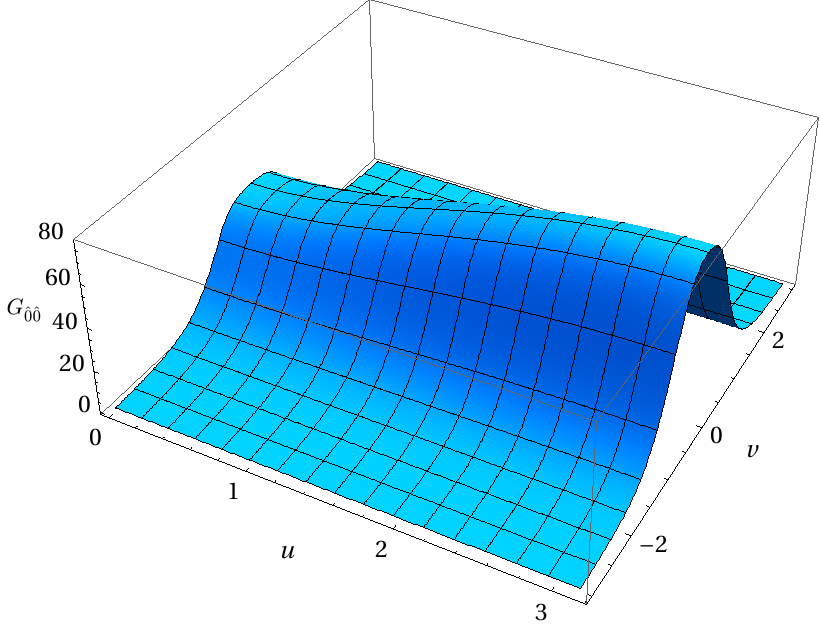}
\caption{$G_{\hat{0}\hat{0}}$ is always positive, indicating the need for positive mass-energy to support the catenary wormhole. The region near $u=\pi$ has a higher Ricci scalar, implying a denser area near the outermost part of the wormhole.}
\label{fig1}
\end{figure}

\newpage
\begin{figure}[h]
\centering
\includegraphics[width=10cm]{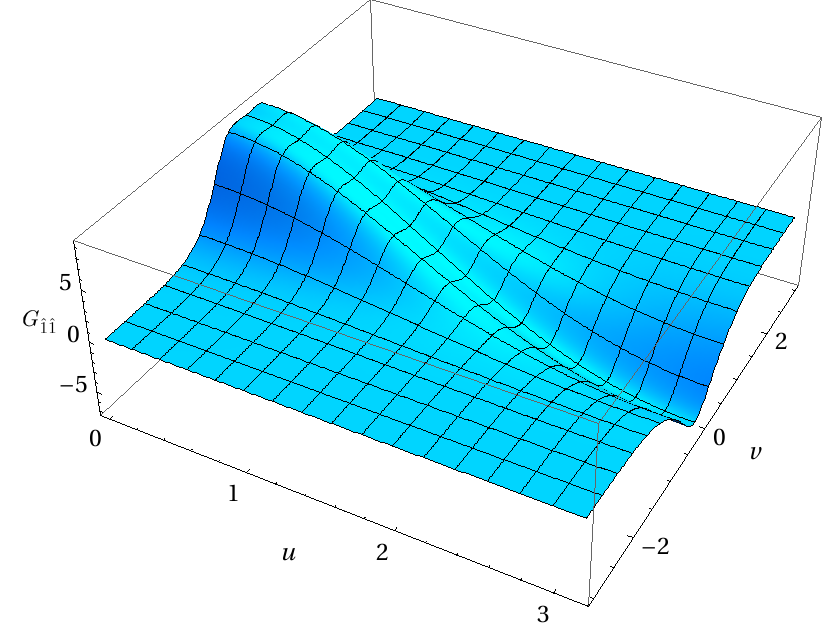}
\caption{$G_{\hat{1}\hat{1}}$ is negative around the outer part of the wormhole ($u=\pi$), but is positive around the inner part ($u=0$).}
\label{fig2}
\end{figure}

\newpage
\begin{figure}[h]
\centering
\includegraphics[width=10cm]{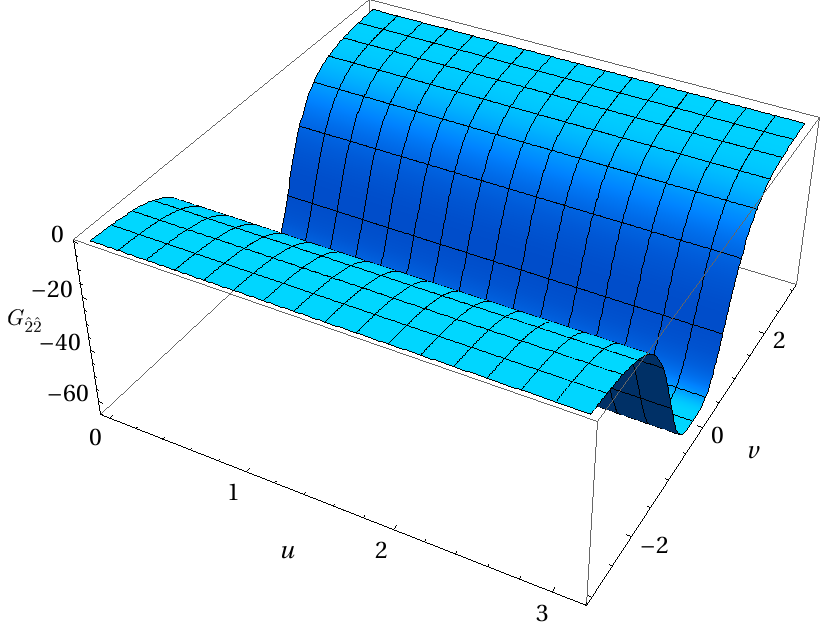}
\caption{The pressure along the $v$-direction is always negative.}
\label{fig3}
\end{figure}

\newpage
\begin{figure}[h]
\centering
\includegraphics[width=10cm]{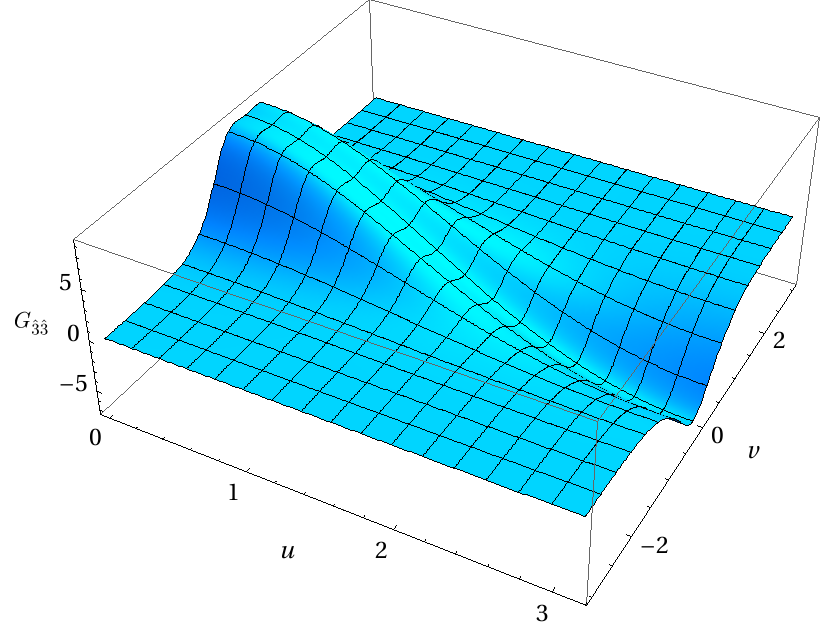}
\caption{$G_{\hat{3}\hat{3}}$ has a variation over $u$ and $v$ that is similar to $G_{\hat{1}\hat{1}}$.}
\label{fig4}
\end{figure}

\newpage
\begin{figure}[h]
\centering
\includegraphics[width=10cm]{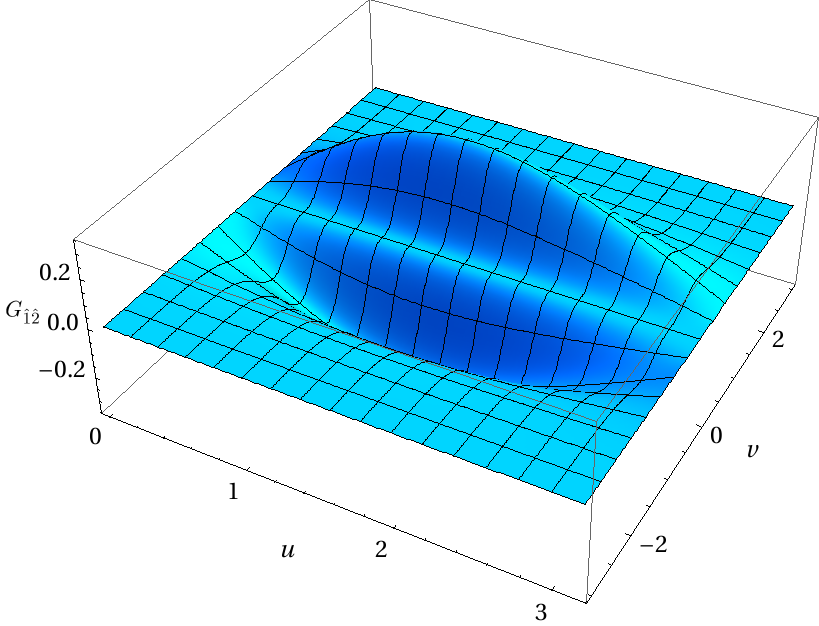}
\caption{There is a relatively small shear stress $G_{\hat{1}\hat{2}}=G_{\hat{2}\hat{1}}$ compared to the other components of $G_{\hat{\mu}\hat{\nu}}$. Note that $G_{\hat{1}\hat{2}}=G_{\hat{2}\hat{1}}$ is zero when $u=0$ and $u=\pi$. This shear stress is also absent at the throat $v=0$.}
\label{fig5}
\end{figure}

\newpage
\begin{figure}[h]
\centering
\includegraphics[width=11.5cm]{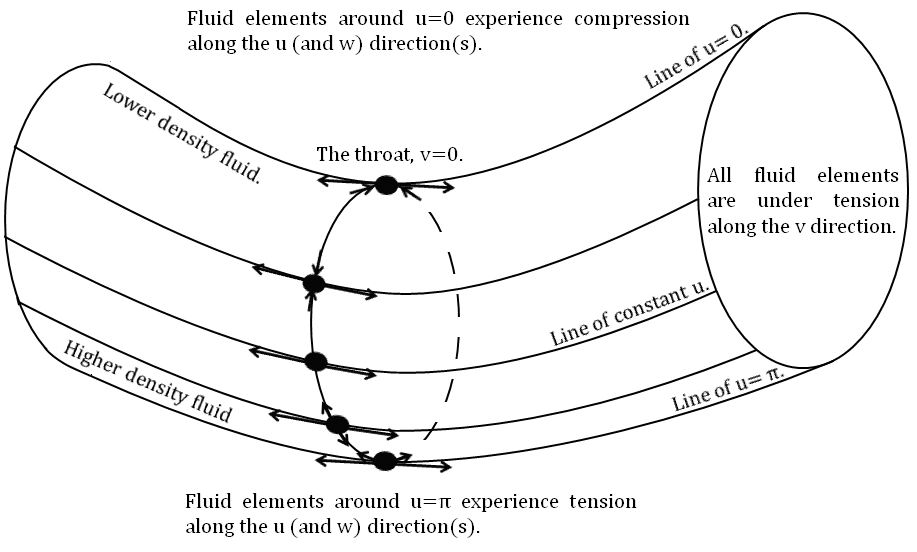}
\caption{The compression and tension experienced by fluid elements at the throat of the catenary wormhole, as well as the relative density of the fluid elements. The $w$-coordinate is suppresed.}
\label{fig6}
\end{figure}

\newpage
\begin{figure}[h]
\centering
\includegraphics[width=10cm]{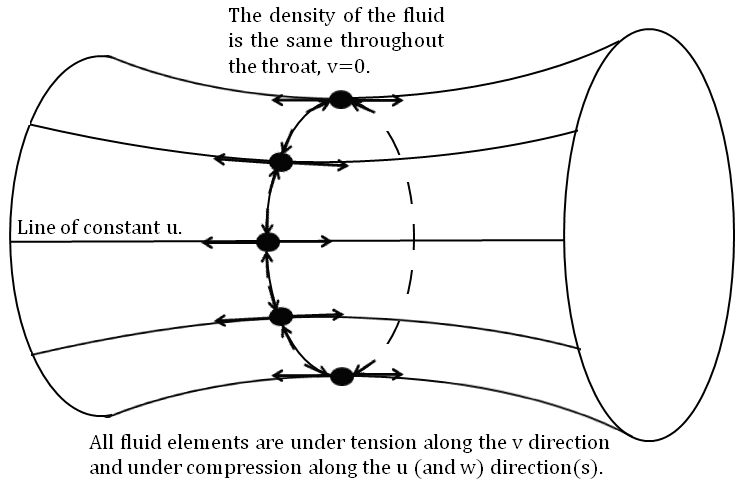}
\caption{The corresponding fluid elements at the throat of a spherically symmetric wormhole.}
\label{fig7}
\end{figure}

\newpage
\begin{figure}[h]
\centering
\includegraphics[width=10cm]{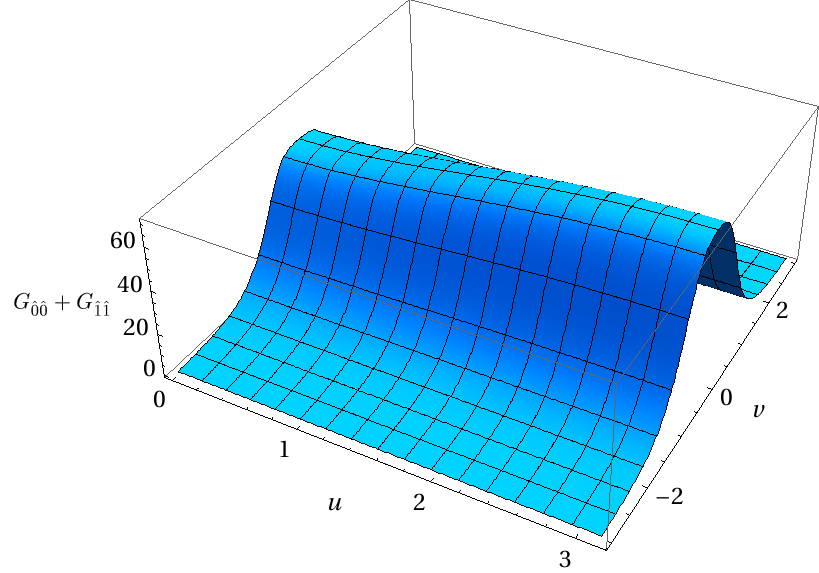}
\caption{Graph of $G_{\hat{0}\hat{0}}+G_{\hat{1}\hat{1}}$ as a function of $u$ and $v$.}
\label{fig8}
\end{figure}

\newpage
\begin{figure}[h]
\centering
\includegraphics[width=10cm]{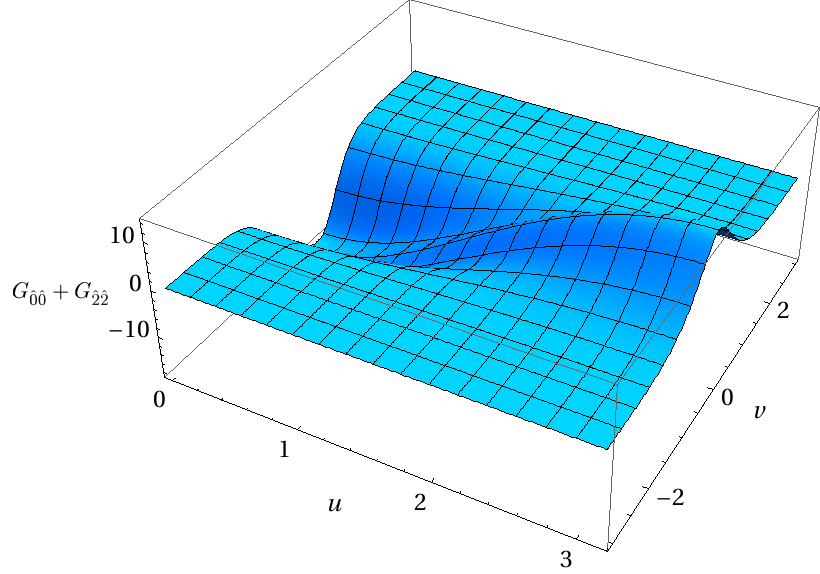}
\caption{Graph of $G_{\hat{0}\hat{0}}+G_{\hat{2}\hat{2}}$ as a function of $u$ and $v$. The null energy condition is violated at the inner region of the catenary wormhole near $u=0$, since it is negative. Nevertheless, this has a positive value in the outer region near $u=\pi$.}
\label{fig9}
\end{figure}

\newpage
\begin{figure}[h]
\centering
\includegraphics[width=10cm]{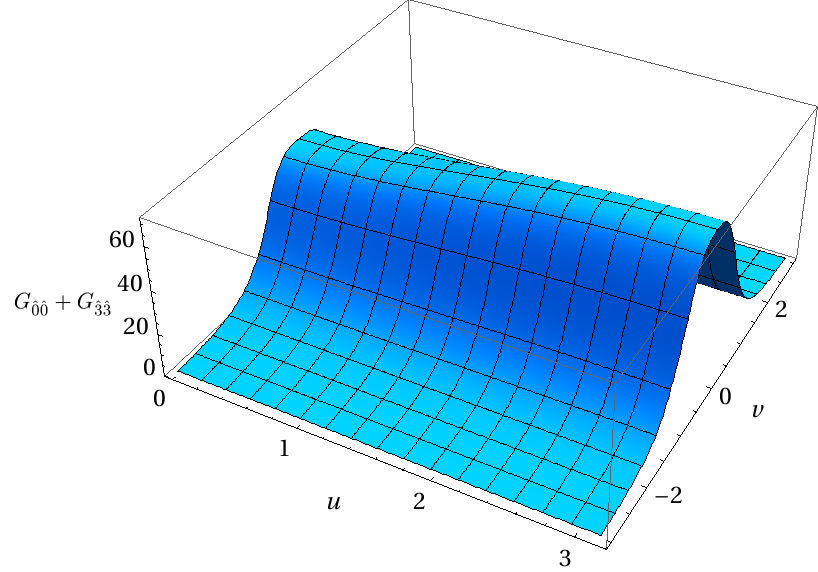}
\caption{Graph of $G_{\hat{0}\hat{0}}+G_{\hat{3}\hat{3}}$ as a function of $u$ and $v$.}
\label{fig10}
\end{figure}

\newpage
\begin{figure}[h]
\centering
\includegraphics[width=10cm]{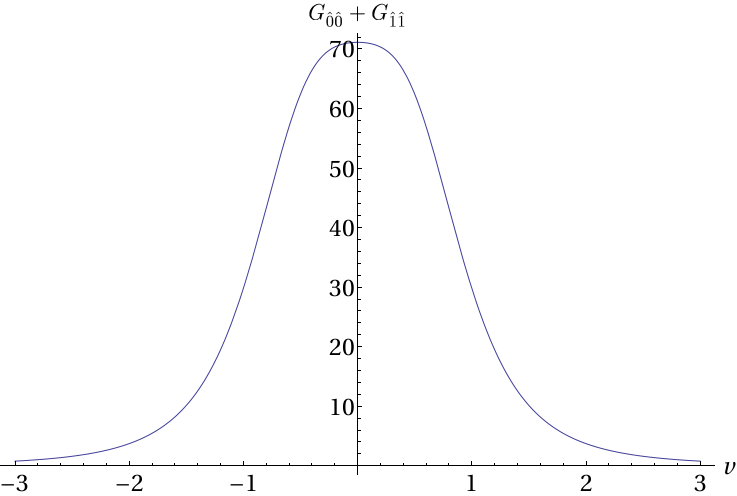}
\caption{Graph of $G_{\hat{0}\hat{0}}+G_{\hat{1}\hat{1}}$ as a function of $v$ at $u=\pi$.}
\label{fig11}
\end{figure}

\newpage
\begin{figure}[h]
\centering
\includegraphics[width=10cm]{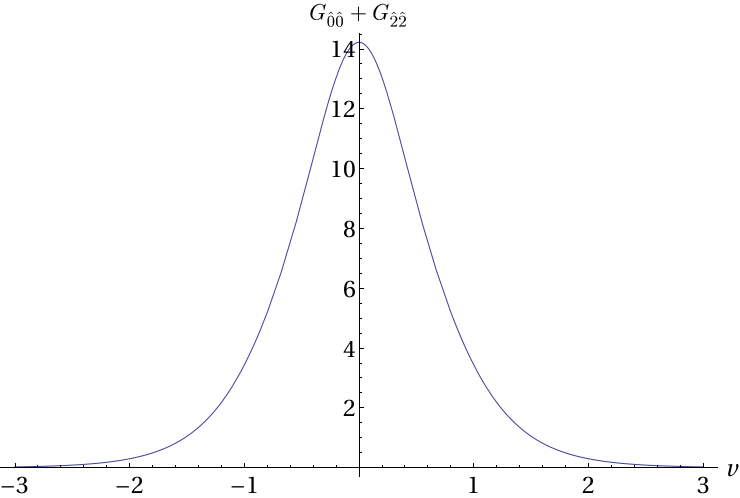}
\caption{Graph of $G_{\hat{0}\hat{0}}+G_{\hat{2}\hat{2}}$ as a function of $v$ at $u=\pi$.}
\label{fig12}
\end{figure}

\newpage
\begin{figure}[h]
\centering
\includegraphics[width=10cm]{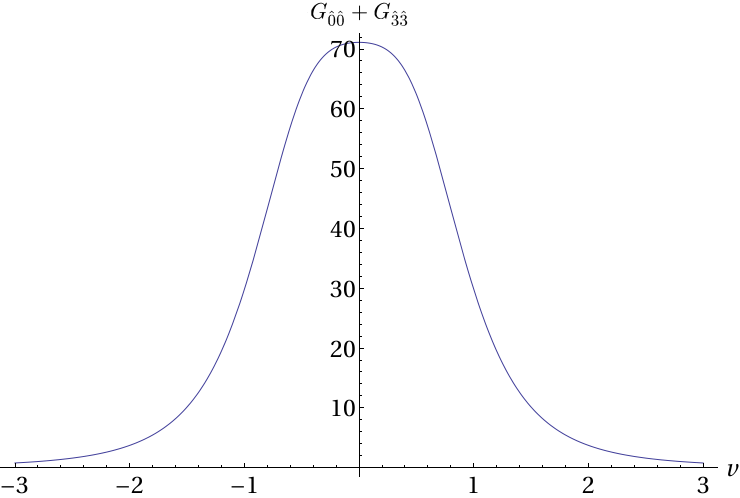}
\caption{Graph of $G_{\hat{0}\hat{0}}+G_{\hat{3}\hat{3}}$ as a function of $v$ at $u=\pi$. Figs. 11-13 conclude that there exists a safe geodesic through the catenary wormhole where $u=\pi$, corresponding to the outermost part of the wormhole.}
\label{fig13}
\end{figure}

\end{document}